\documentclass[useAMS,usegraphicx,usenatbib]{mn2e}
\usepackage{times}

\begin{document}

\title[Cosmological constraints]{Current cosmological constraints on the curvature, dark energy and modified gravity}

\author[Y.G. Gong et al.]{Yungui Gong,$^1$\thanks{gongyg@cqupt.edu.cn} Xiao-ming Zhu,$^1$ and Zong-Hong Zhu,$^2$
\thanks{zhuzh@bnu.edu.cn}\\
$^1$College of Mathematics and Physics, Chongqing University of Posts and Telecommunications,
Chongqing 400065, China\\
$^2$Department of Astronomy, Beijing Normal university,
Beijing 100875, China}
\maketitle

\begin{abstract}
We apply the Union2 compilation of 557 supernova Ia data, the baryon acoustic oscillation measurement of distance,
the cosmic microwave background radiation data from the seven year Wilkinson Microwave Anisotropy Probe,
and the Hubble parameter data to study the geometry of the Universe and the
property of dark energy by using models and
parametrizations with different high redshift behaviours of $w(z)$. We find that $\Lambda$CDM model is
consistent with current data, that the Dvali-Gabadadze-Porrati
model is excluded by the data at more than $3\sigma$ level, that the Universe is almost flat,
and that the current data is unable to distinguish models with different behaviours of $w(z)$ at high redshift.
We also add the growth factor data to constrain the growth index of Dvali-Gabadadze-Porrati model and
find that it is more than $1\sigma$ away from its theoretical value.
\end{abstract}



\section{Introduction}

Even since the discovery of the accelerated expansion of the Universe in 1998 \citep{acc1,acc2}, many
efforts have been made to confirm and understand this phenomenon of acceleration. For the explanation
of the acceleration, there are three different approaches. The first method introduces a new
exotic form of matter with negative pressure, dubbed as dark energy to drive the Universe to accelerate.
The cosmological constant is the simplest candidate of dark energy which is also consistent
with observations, but at odds with quantum field theory. The second method modifies
the theory of gravity known as general relativity at the cosmological scale, such as
the Dvali-Gabadadze-Porrati (DGP) model \citep{dgp}. The third approach takes the view that the
Universe is inhomogeneous. In this paper, we focus on dark energy and DGP models.

The only effect of dark energy we know is through gravitational interaction; this makes
it difficult to understand the physical nature of dark energy.  In particular,
the question whether dark energy is the cosmological constant remains unanswered.
Recently, it was claimed that
the flat $\Lambda$CDM model is inconsistent with observations at
more than $1\sigma$ level \citep{huang,star,cai}.
\citet{star} suggested that the cosmic acceleration is slowing down from $z\sim 0.3$. In \citet{huang},
it was claimed that dark energy suddenly emerged at redshift $z\sim 0.3$.
\citet{cai} found possible oscillating behaviour of dark energy.
However, no evidence for dark energy
dynamics was found  in other studies \citep{lampeitl,corray9,gong10a,gong10b,pan}.
The difference between the conclusions drawn in \citet{star} and \citet{gong10a}
lies in the Baryon Acoustic Oscillation (BAO) data used in their analysis. \citet{star}
employs the ratio $D_V(0.35)/D_V(0.2)$ of the effective distance $D_V(z)$ at two redshifts,
while \citet{gong10a} applies the BAO $A$ parameter given by \citet{baoa}.
The tension between BAO measurement and higher redshift type Ia supernova (SN Ia) was noticed in \cite{percival07},
and the tension was lessened in \citet{wjp} due to revised error analysis, different methodology adopted
and more data.

It was argued that the systematics in
different data sets heavily affected the fitting results \citep{consta,sollerman,gong10b,sdss2}.
The Constitution compilation found that the scatter at high redshift is higher for SALT and SALT2 fitters, and
SALT2 poorly fits the nearby U-band light curves \citep{consta}. However,
it was found that SALT2 performs better than SALT and MLCS2k2 judged by the scatter
around the best-fitting luminosity distance relationship  in \cite{conley} and \cite{union2}. Because
MLCS2k2 training is mainly based on the observation of nearby SN Ia, and because the observations made in the
observer-frame U-band are contaminated with a high level of uncertainty due to atmospheric variations,
so MLCS2k2 is less accurate at predicting the rest-frame U-band using high redshift SN Ia \citep{union2,sdss2}.
The Union2 data applies the SALT2 light curve fitter because it addresses the problem by including high
redshift data where the rest-frame U-band is observed at redder wavelengths \citep{union2}.
In this paper, we combine the Union2 sample of 557 SN Ia data with systematic errors \citep{union2},
the BAO distance ratios $r_s(z_d)/D_V(z)$ between the comoving sound horizon at the baryon drag epoch $r_s(z_d)$
and the effective distance $D_V(z)$ at $z=0.2$ and $z=0.35$ \citep{wjp}, the radial BAO measurements
at $z=0.24$ and $z=0.43$ \citep{eg}, the seven-year Wilkinson Microwave
Anisotropy Probe (WMAP7) data \citep{wmap7},
and the Hubble parameter $H(z)$ data \citep{hz2,hz1} to probe the geometry of the Universe and
the nature of dark energy with different models.

The paper is organised as follows. In section 2, we present the SN Ia data \citep{union2},
the BAO data \citep{eg,wjp}, the WMAP7 data \citep{wmap7}, the $H(z)$ data \citep{hz2,hz1}, and
the growth factor data \citep{viel1,viel2,mcdonald,tegmark,ross,angela,guzzo,blake10}, and
all the formulae related to these data.
In section 3, we list all the models and the fitting results,
and conclusions are given in section 4.

\section{Observational data}
The Union2 SN Ia data consist of the low-$z$ SN Ia data observed at the F.L. Whipple  observatory of
the Harvard-Smithsonian centre for astrophysics \citep{consta},
the intermediate-$z$ data observed during the first season of the Sloan Digital Sky
Survey (SDSS)-II supernova survey \citep{sdss2}, and the high-$z$ data from the Union compilation \citep{union}.
The Union2 SN Ia data used the SALT2 light-curve fitter because it performs better
than both SALT and MLCS2k2 when judged by the scatter around the best-fitting luminosity
distance relationship \citep{union2}.
To use the 557 Union2 SN Ia data \citep{union2},
we minimize
\begin{equation}
\label{chi}
\chi^2=\sum_{i,j=1}^{557}[\mu(z_i)-\mu_{obs}(z_i)]C^{-1}_{sn}(z_i,z_j)[\mu(z_j)-\mu_{obs}(z_j)],
\end{equation}
where the extinction-corrected distance modulus
$\mu(z)=5\log_{10}[d_L(z)/{\rm Mpc}]+25$, $C_{sn}(z_i,z_j)$ is the covariant matrix which includes
the systematical errors for the SN Ia data \citep{union2}; the covariant matrix is available online\footnote{http://supernova.lbl.gov/Union/}.
The luminosity distance $d_L(z)$
is
\begin{equation}
\label{lum} d_L(z)=\frac{1+z}{H_0\sqrt{|\Omega_{k}|}}\, S_k\!\left[\sqrt{|\Omega_{k}|}\int_0^z \frac{dx}{E(x)}\right],
\end{equation}
the dimensionless Hubble parameter $E(z)=H(z)/H_0$;
and $S_k(x)$ is defined as $x$, $\sin(x)$ or $\sinh(x)$ for $k=0$, +1, or -1, respectively.
Due to the arbitrary normalization of the luminosity distance, the nuisance parameter
$h=H_0/$(100 km s$^{-1}$Mpc$^{-1}$) in the SN Ia data is not the observed Hubble constant.
So we marginalize the nuisance parameter $h$ with a flat prior,
after the marginalization, we get \citep{gong08}
\begin{eqnarray}
\begin{array}{ll}
\label{chi1} \chi^2_{sn}(\mathbf{p})=&\sum_{ij}\alpha_i C^{-1}_{sn}(z_i,z_j)\alpha_j\\
&-\frac{\left[\sum_{ij}\alpha_i C^{-1}_{sn}(z_i,z_j)-\ln 10/5\right]^2}{\sum_{ij} C^{-1}_{sn}(z_i,z_j)}\\
&-2\ln\left(\frac{\ln 10}{5}\sqrt{\frac{2{\rm \pi}}{\sum_{ij} C^{-1}_{sn}(z_i,z_j)}}\right),
\end{array}
\end{eqnarray}
where $\alpha_i=\mu_{obs}(z_i)-25-5\log_{10}[H_0 d_L(z_i)]$.

In addition to the Union2 SN Ia data, we  use the BAO distance measurement
from the oscillations in the distribution of galaxies.
The BAO is due to the sound waves in the plasma of the early Universe and the wavelength of the BAO
is related to the comoving sound horizon at the baryon drag epoch. The distance at the redshift $z=0.2$
was measured in the clustering of the combined 2dF Galaxy Redshift Survey (2dFGRS) and SDSS main galaxy samples,
and the distance at the redshift $z=0.35$ was measured in the clustering of the SDSS luminous red galaxies.
From the BAO observation of the galaxy power spectra, \cite{wjp} measured the distance ratio,
\begin{equation}
\label{dz} d_{z}= \frac{r_{s}(z_{d})}{D_{V}(z)}
\end{equation}
at two redshifts $z=0.2$ and $z=0.35$ to be $d_{0.2}^{obs}=0.1905\pm 0.0061$,
and $d_{0.35}^{obs}=0.1097\pm 0.0036$ (hereafter Bao2).
Here the effective distance is
\begin{equation}
\label{dvdef}
D_V(z)=\left[\frac{d_L^2(z)}{(1+z)^2}\frac{z}{H(z)}\right]^{1/3},
\end{equation}
$z_d$ is the drag redshift defined in \cite{dw},
the comoving sound horizon is
\begin{equation}
\label{rshordef} r_s(z)=\int_z^\infty \frac{c_s(x)dx}{E(x)},
\end{equation}
where the sound speed $c_s(z)=1/\sqrt{3[1+\bar{R_b}/(1+z)}]$, and
$\bar{R_b}=3\Omega_b h^2/(4\times2.469\times10^{-5})$.
To use this BAO data, we calculate
\begin{equation}
\label{baochi2} \chi^2_{Bao2}(\mathbf{p},\Omega_b h^2, h)=\sum_{i,j=1}^{2}\Delta d_i
C_{Bao}^{-1}(d_i,d_j)\Delta d_j,
\end{equation}
where $d_i=(d_{z=0.2},d_{z=0.35})$, $\Delta d_i=d_i-d_i^{obs}$ and
the covariance matrix $C_{Bao}(d_i,d_j)$ for the two parameters $d_{0.2}$ and $d_{0.35}$
is taken from equation (5) in \cite{wjp}. Besides the model parameters $\mathbf{p}$, we need
to add two more nuisance parameters $\Omega_b h^2$ and $\Omega_m h^2$ when we use the BAO data.

From the measurement of the radial (line-of-sight) BAO scale in the
galaxy power spectra, the cosmological parameters may be determined
from the measured values of
\begin{equation}
\label{deltaz} \Delta z_{Bao}(z)=\frac{H(z)r_{s}(z_{d})}{c}
\end{equation}
at two redshifts $z=0.24$ and $z=0.43$, which are
$\Delta z_{Bao}(z=0.24)=0.0407\pm 0.0011$ and $\Delta z_{Bao}(z=0.43)=0.0442\pm0.0015$ (hereafter Baoz), respectively \citep{eg}.
Therefore, we add $\chi^{2}$ with
\begin{eqnarray}
\label{baochi3}
\begin{array}{ll}
\chi^2_{Baoz}(\mathbf{p},\Omega_b h^2, h)=&\left(\frac{\Delta
z_{Bao}(0.24)-0.0407}{0.0011}\right)^2 \\
&+\left(\frac{\Delta z_{Bao}(0.43)-0.0442}{0.0015}\right)^2.
\end{array}
\end{eqnarray}
When we add this BAO data to the fitting, we also need to add the
nuisance parameters $\Omega_b h^2$ and $\Omega_m h^2$. In \cite{gong10b},
it was found that the Baoz data is consistent with the Bao2 data, and the constraints
on the model parameters get improved with the addition of the Baoz data.

Because both the SN Ia and the BAO data measure the distance up to redshit $z<2$,
we need to add distance data at higher redshift $z>10$ to better constrain
the property of dark energy, so we use the WMAP7 data.
When the full WMAP7 data are applied, we need to add some more parameters
which depend on inflationary models, and this will limit our ability to constrain dark energy models.
So we only use the WMAP7 measurements of the derived quantities such as the shift parameter
$R(z^{*})$ and the acoustic scale $l_A(z^{*})$ at the decoupling redshift, and the decoupling redshift $z^{*}$.
Then we add the following term to $\chi^2$,
\begin{equation}
\label{cmbchi} \chi^2_{CMB}(\mathbf{p},\Omega_b h^2, h)=\sum_{i,j=1}^{3}\Delta x_i C_{CMB}^{-1}(x_i,x_j)\Delta x_j,
\end{equation}
where the three parameters $x_i=[R(z^{*}),\ l_A(z^{*}),\ z^{*}]$, $\Delta
x_i=x_i-x_i^{obs}$ and the covariance matrix $C_{CMB}(x_i,x_j)$
for the three parameters is taken from Table 10 in \cite{wmap7}. The shift parameter $R$ is
expressed as
\begin{equation}
\label{shift}
R(z^{*})=\frac{\sqrt{\Omega_{m}}}{\sqrt{|\Omega_{k}|}}S_k\!\left(\sqrt{|\Omega_{k}|}\int_0^{z^{*}}\frac{dz}{E(z)}\right)
=1.725\pm 0.018.
\end{equation}
 The acoustic scale $l_A$ is
\begin{equation}
\label{ladefeq} l_A(z^{*})=\frac{{\rm \pi}
d_L(z^{*})}{(1+z^{*})r_s(z^{*})}=302.09\pm0.76,
\end{equation}
and $z^{*}$ is the decoupling redshift with the parametrization defined in  \cite{hs}.
We also need to add the parameters $\Omega_b h^2$ and $\Omega_m h^2$ to the parameter space when we employ the WMAP7 data.

The SN Ia, BAO and WMAP7 data measured the distance which depends on
the double integration of the equation of state parameter $w(z)$, the process of double
integration smoothes out the variation of equation of state parameter $w(z)$ of dark energy.
To alleviate the double integration, we also apply the measurements of the Hubble parameter $H(z)$
which depends on $\Omega_{DE}$ directly and detects the variation of $w(z)$ better than the distance scales.
Furthermore, the addition of the $H(z)$ data can better constrain $w(z)$ at high redshift \citep{gong10a}.
In this paper, we use the $H(z)$ data at 11 different redshifts obtained from the differential ages of
red-envelope galaxies in \cite{hz1}, and three more Hubble parameter data
$H(z=0.24)=76.69\pm2.32$, $H(z=0.34)=83.8\pm2.96$ and
$H(z=0.43)=86.45\pm3.27$, determined by \cite{hz2}. So we add these $H(z)$ data to $\chi^2$,
\begin{equation}
\label{hzchi}
\chi^2_H(\mathbf{p}, h)=\sum_{i=1}^{14}\frac{[H(z_i)-H_{obs}(z_i)]^2}{\sigma_{hi}^2},
\end{equation}
where $\sigma_{hi}$ is the $1\sigma$ uncertainty in the $H(z)$
data.
Basically,
The model parameters $\mathbf{p}$ are determined by minimizing
\begin{equation}
\label{chi2min}
\chi^2=\chi^2_{sn}+\chi^2_{Bao2}+\chi^2_{Baoz}+\chi^2_{CMB}+\chi^2_H.
\end{equation}
In addition, we add the prior $h=0.742\pm 0.036$ \citep{riess09}.

In order to distinguish
the modified gravity such as DGP model from dark energy models, we approximate the growth factor
with $f(z)=\Omega_{m}^\gamma+(\gamma-4/7)\Omega_{k}$ \citep{gong09}, then we use the growth factor data $f(z)$
obtained from the measurement of the redshift distortion to constrain the growth index $\gamma$ of the models
\citep{viel1,viel2,mcdonald,tegmark,ross,angela,gongprd08,guzzo,blake10,ishak}.
So we calculate
\begin{equation}
\label{chi2fz}
\chi^2_f(\mathbf{p},\gamma)=\sum_{i=1}^{15}\frac{[f(z_i)-\Omega_m^\gamma(z_i)-(\gamma-4/7)\Omega_k(z_i)]^2}{\sigma_{f_i}^2}.
\end{equation}

The likelihood for the parameters $\mathbf{p}$ in the
model and the nuisance parameters is
computed using the Monte Carlo Markov Chain (MCMC) method.
The MCMC method
randomly chooses values for the above parameters $\mathbf{p}$, evaluates $\chi^2$
and determines whether to accept or reject the set of parameters $\mathbf{p}$
using the Metropolis-Hastings algorithm. The set of parameters that
are accepted to the chain forms a new starting point for the next
process, and the process is repeated for a sufficient number of
steps until the required convergence is reached. Our MCMC code is
based on the publicly available package {\sc cosmomc} \citep{cosmomc,gong08}.

After fitting the observational data to different dark energy models, we apply the $Om$ diagnostic \citep{omz}
to detect the deviation from the $\Lambda$CDM model. For a flat universe \citep{omz},
\begin{equation}
\label{omz}
Om(z)=\frac{E^2(z)-1}{(1+z)^3-1}.
\end{equation}
For the $\Lambda$CDM model, $Om(z)=\Omega_m$ is a constant which is independent of the value of $\Omega_m$.
Because of this property, $Om$ diagnostic is less sensitive to observational errors
than the equation of state parameter $w(z)$ does.
On the other hand, the bigger the value of $Om(z)$, the bigger the value of $w(z)$.

\section{Cosmological models}

\subsection{$q_1-q_2$ parametrization}
To understand the current accelerating expansion, we parametrize the deceleration parameter $q(z)$
with a simple two-parameter function \citep{gong07a},
\begin{equation}
\label{qmod2}
q(z)=\frac{1}{2}+\frac{q_1 z+q_2}{(1+z)^2}.
\end{equation}
In this parametrization, we have only two parameters $\mathbf{p}=(q_1,\ q_2)$.
The parameter $q_2=q(z=0)-1/2$, and $q(z)=1/2$ at high redshift which represents the matter dominated epoch.
In principle, this parametrization does not involve $\Omega_m$ and $\Omega_k$,
but the comoving distance depends on the geometry of the Universe through the function $S_k$,
for simplicity, we consider
the flat case $\Omega_k=0$ for this model. Although the flat assumption of $\Omega_k=0$ may induce large
error in the estimation of cosmological parameters due to the degeneracy among $\Omega_m$, $\Omega_k$ and
$w$ \citep{clarkson}, but for this model, the only effect of $\Omega_k$
is through $S_k$, and $S_k(x)\approx x$ when $\Omega_k$ is small, so the impact of the flat assumption is small.
The dimensionless Hubble parameter is
\begin{eqnarray}
\label{hubsl2}
\begin{array}{ll}
E(z)&=\exp\left[\int_0^z[1+q(u)]d\ln(1+u)\right]\\
&=(1+z)^{3/2}\exp\left[\frac{q_2}{2}+\frac{q_1 z^2-q_2}{2(1+z)^2}\right].
\end{array}
\end{eqnarray}
When $z\gg 1$, $E^2(z)\approx (1+z)^3\exp(q_1+q_2)$, so we can think
$q_1+q_2=\ln\Omega_m$.
To account for the radiation-dominated Universe, we take the above $E(z)$ as
the approximation for the matter and dark energy only, so we use the following Hubble parameter to approximate
the whole expansion history of the Universe,
\begin{equation}
\label{hubsl3}
E^2(z)=\Omega_r(1+z)^4+(1+z)^3\exp\left[q_2+\frac{q_1 z^2-q_2}{(1+z)^2}\right],
\end{equation}
where the current radiation component $\Omega_{r}=4.1736\times 10^{-5}h^{-2}$ \citep{wmap7}.
Fitting the model to the combined SN Ia, Bao2, Baoz, WMAP7 and $H(z)$ data, we get the marginalized $1\sigma$
constraints, $q_1=0.07\pm 0.11$ and $q_2=-1.43\pm 0.09$ with $\chi^2=542.6$.
In terms of $q_0=q(z=0)$, we find that $q_0<-0.62$ at $3\sigma$ confidence level, so the evidence of current acceleration is very strong.
Furthermore, we find that the $3\sigma$ constraint on $\Omega_m$ is $\Omega_m=0.257_{-0.035}^{+0.044}$. The contour plot
for $\Omega_m$ and $q_0$ is shown in Fig. \ref{u2cont}(d).

At a low redshift, the radiation is negligible, so $Om(z)$ in this model is
\begin{equation}
\label{qomz}
Om(z)=\frac{(1+z)^3\exp[q_2+(q_1 z^2-q_2)/(1+z)^2]-1}{(1+z)^3-1}.
\end{equation}
By using the best-fitting values of $q_1$ and $q_2$, we reconstruct $Om(z)$ and the results are plotted in Fig. \ref{u2omz}(d).

\subsection{Piecewise parametrization of $q(z)$}
To further study the evolution of the deceleration parameter $q(z)$,
we use the more model-independent piecewise parametrizations.
We group the data into four bins so that the number of SN Ia in each bin times the width of each bin is around 30, i.e.,
$N\Delta z\sim 30$ and $N=4$. The four bins are $z_1=0.1$, $z_2=0.4$, $z_3=0.7$, $z_4=1.8$ and $z_5$ extends beyond 1089.
For the redshift in the range $z_{i-1}\le z<z_i$, the deceleration parameter $q(z)$ is a constant $q_i$, $q(z)=q_i$.
Take $z_0=0$, then for $z_{i-1}\le z<z_i$, we get
\begin{equation}
\label{qzcez}
E(z)=(1+z)^{1+q_N}\prod_{i=1}^{N}(1+z_{i-1})^{q_{i-1}-q_i}.
\end{equation}
In this model, we have four parameters $\mathbf{p}=(q_1,\ q_2,\ q_3,\ q_4)$.
In general, for the piecewise parametrisation, the parameters such as $q_i$ in different bins
are correlated and their errors depend upon each other. We follow
\cite{huterer05} to transform the covariance matrix of
$q_i$s to decorrelate the error estimate. Explicitly, the
transformation is
\begin{equation}
\label{decorr}
{\mathcal Q}_i=\sum_j T_{ij}q_j,
\end{equation}
where the transformation matrix $T=V^T\Gamma^{-1/2}V$, the orthogonal matrix
$V$ diagonalizes the covariance matrix $C$ of $q_i$ and $\Gamma$ is the diagonalized matrix of $C$.
For a given $i$, $T_{ij}$ can be thought of as weights for each $q_j$ in the transformation from
$q_i$ to ${\mathcal Q}_i$. We are free to rescale each ${\mathcal Q}_i$ without changing the
diagonality of the correlation matrix, so we then multiply both sides of the equation above by an
amount such that the sum of the weights $\sum_j T_{ij}$ is equal to 1. This allows for easy
interpretation of the weights as a kind of discretized window function. Now the transformation matrix
element is $T_{ij}/\sum_k T_{ik}$ and the covariance matrix of the uncorrelated parameters is not the
identity matrix. The $i$-th diagonal matrix element becomes $(\sum_j T_{ij})^{-2}$. In other words,
the error of the uncorrelated parameters ${\mathcal Q}_i$ is $\sigma_i=1/\sum_j T_{ij}$.
Fitting the model to the combined SN Ia, Bao2, Baoz, WMAP7 and $H(z)$ data, we get the constraints
on the uncorrelated parameters $\mathcal{Q}_i$ and the result is shown in Fig. \ref{qzunc}.

\subsection{$\Lambda$CDM model}
For the cosmological constant, the equation of state parameter $w=p/\rho=-1$,
and the energy density $\rho_\Lambda$ is a constant.
For a curved $\Lambda$CDM model, the curvature term $\Omega_k\neq 0$,
Friedmann equation is
\begin{equation}
\label{elcdm}
E^2(z)=\Omega_{k}(1+z)^2+\Omega_{m}(1+z)^3
 +\Omega_{r}(1+z)^4+\Omega_{\Lambda}.
\end{equation}
At low redshits, the contribution of the radiation term is negligible.
We have two parameters $\mathbf{p}=(\Omega_m,\ \Omega_k)$
and one nuisance parameter $h$ in this model.
By fitting the $\Lambda$CDM model to the combined SN Ia, Bao2, Baoz, WMAP7 and $H(z)$ data, we get
the marginalized $1\sigma$ constraints, $\Omega_m=0.272^{+0.013}_{-0.011}$
and $\Omega_k=0.002\pm 0.004$ with $\chi^2=541.2$.
The contours of $\Omega_m$ and $\Omega_k$
are plotted in Fig. \ref{u2omkcont}(a).
By fitting the model to observational data combined with the growth factor
data, the marginalized $1\sigma$ constraints are, $\Omega_m=0.272^{+0.013}_{-0.01}$,
$\Omega_k=0.002\pm 0.004$ and $\gamma=0.56^{+0.14}_{-0.09}$ with $\chi^2=546.3$.

\subsection{DGP model}
In this model, gravity appears four dimensional at short distances while modified at
large distances \citep{dgp}. The model is motivated by brane cosmology in which our
Universe is a three-brane embedded in a five dimensional spacetime. The Friedmann equation is modified as
\begin{equation}
\label{dgpez}
E^2(z)=\Omega_{k}(1+z)^2+[
\Omega_{d}+\sqrt{\Omega_{m} (1+z)^3+\Omega_{d}^2}\,]^2,
\end{equation}
where $\Omega_{d}=(1-\Omega_{m}-\Omega_{k})/2\sqrt{1-\Omega_{k}}$. If we take the point of view
that Friedmann equation is not modified and the extra term in equation (\ref{dgpez}) is dark energy, then the equivalent dark energy
equation of state parameter $w(z)$ for the DGP model is
\begin{equation}
\label{dgpwz}
w(z)=-\frac{\Omega_{m}(1+z)^3+2\Omega_{d}[\sqrt{\Omega_{m}(1+z)^3+\Omega_{d}^2}+\Omega_{d}]}{2
[\Omega_{m}(1+z)^3+\Omega_{d}^2+\Omega_{d}\sqrt{\Omega_{m}(1+z)^3+\Omega_{d}^2}]}.
\end{equation}
when $z\gg 1$, $w(z)\sim -1/2$ and $w(z=0)=-(1-\Omega_{k})/(1+\Omega_{m}-\Omega_{k})$. Since $\Omega_{k}$
is very small, $w(z)>-1$ for the DGP model.

By fitting the DGP model to the combined SN Ia, Bao2, Baoz, WMAP7 and $H(z)$ data, the marginalized $1\sigma$
constraints are $\Omega_m=0.288^{+0.015}_{-0.011}$ and $\Omega_k=0.019\pm 0.005$ with $\chi^2=561.6$.
By fitting the DGP model to the combined SN Ia, Bao2, Baoz, WMAP7, $H(z)$, and $f(z)$ data, we get
the marginalized $1\sigma$ estimations $\Omega_m=0.290^{+0.014}_{-0.012}$, $\Omega_k=0.019\pm 0.005$,
and $\gamma=0.46^{+0.12}_{-0.08}$ with $\chi^2=567.5$.

\subsection{CPL parametrization}

For the Chevallier-Polarski-Linder (CPL) parametrization \citep{cpl1,cpl2}, the equation of state parameter is
\begin{equation}
\label{lind}
w(z)=w_0+\frac{w_a z}{1+z},
\end{equation}
so $w(z=0)=w_0$ and $w(z)\sim w_0+w_a$ when $z\gg 1$. The corresponding dimensionless dark energy density is
\begin{equation}
\label{cplde}
\Omega_{DE}(z)=\Omega_x (1+z)^{3(1+w_0+w_a)}{\rm e}^{[-3w_az/(1+z)]},
\end{equation}
where $\Omega_x=1-\Omega_{m}-\Omega_r-\Omega_{k}$. In this model, we have four model parameters $\mathbf{p}=(\Omega_{m},\ \Omega_{k},  \ w_0, \ w_a)$.
Fitting the model to the combined SN Ia, Bao2, Baoz, WMAP7 and $H(z)$ data, we get the marginalized
$1\sigma$ constraints, $\Omega_m=0.265^{+0.019}_{-0.009}$, $\Omega_k=0.008^{+0.005}_{-0.011}$,
$w_0=-1.16^{+0.26}_{-0.06}$, and $w_a=0.69^{+0.24}_{-1.41}$ with $\chi^2=540.5$.
The contours of $w_0$ and $w_a$
are plotted in Fig. \ref{u2cont}(a), and the contours of $\Omega_m$ and $\Omega_k$
are plotted in Fig. \ref{u2omkcont}(b).

For the flat CPL model, $Om(z)$ becomes
\begin{equation}
\label{cplomz}
Om(z)=\frac{\Omega_m (1+z)^3+\Omega_{DE}(z)-1}{(1+z)^3-1},
\end{equation}
where $\Omega_{DE}(z)$ is defined in equation (\ref{cplde}) with $\Omega_k=0$.
By fitting the combined data to the flat CPL model, we get the marginalized $1\sigma$ constraints,
$\Omega_m=0.267^{+0.019}_{-0.01}$, $w_0=-1.05^{+0.17}_{-0.1}$, and $w_a=0.07^{+0.32}_{-0.88}$
with $\chi^2=541.1$. Using this result, we reconstruct $Om(z)$ with equation (\ref{cplomz}) and
the result is shown in Fig. \ref{u2omz}(a).

\subsection{JBP parametrization}
For the Jassal-Bagla-Padmanabhan (JBP) parametrization \citep{jbp}, the equation
 of state parameter is
\begin{equation}
\label{jbpw} w(z)=w_0+\frac{w_a z}{(1+z)^2},
\end{equation}
so $w(z=0)=w_0$ and $w(z)\sim w_0$ when $z\gg 1$. In this model, the parameter $w_0$
determines the property of the equation of state parameter $w(z)$ at both low and high redshifts.
The corresponding dimensionless dark energy density is then
\begin{equation}
\label{jbpde}
\Omega_{DE}(z)=\Omega_x(1+z)^{3(1+w_0)}{\rm e}^{[3w_az^2/2(1+z)^2]},
\end{equation}
where $\Omega_x=1-\Omega_{m}-\Omega_r-\Omega_{k}$. In this model, we also have four parameters $\mathbf{p}=(\Omega_{m},\ \Omega_{k},  \ w_0, \ w_a)$.
Fitting the model to the combined SN Ia, Bao2, Baoz, WMAP7 and $H(z)$ data, we get the marginalized
$1\sigma$ constraints, $\Omega_m=0.263^{+0.02}_{-0.01}$, $\Omega_k=0.004\pm 0.006$,
$w_0=-1.21^{+0.32}_{-0.18}$, and $w_a=1.29^{+1.35}_{-2.33}$ with $\chi^2=540.6$.
The contours of $\Omega_m$ and $\Omega_k$
are plotted in Fig. \ref{u2omkcont}(c), and the contours of $w_0$ and $w_a$
are plotted in Fig. \ref{u2cont}(b).

For the flat JBP model, $Om(z)$ becomes
\begin{equation}
\label{jbpomz}
Om(z)=\frac{\Omega_m (1+z)^3+\Omega_{DE}(z)-1}{(1+z)^3-1},
\end{equation}
where $\Omega_{DE}(z)$ is defined in equation (\ref{jbpde}) with $\Omega_k=0$.
By fitting the combined data to the flat JBP model, we get the marginalized $1\sigma$ constraints,
$\Omega_m=0.265^{+0.019}_{-0.011}$, $w_0=-1.08^{+0.24}_{-0.19}$, and $w_a=0.32^{+1.01}_{-1.72}$
with $\chi^2=541.0$. Using this result, we reconstruct $Om(z)$ with equation (\ref{jbpomz}) and
the result is shown in Fig. \ref{u2omz}(b).

\subsection{Wetterich parametrization}
\begin{figure}
\includegraphics[width=84mm]{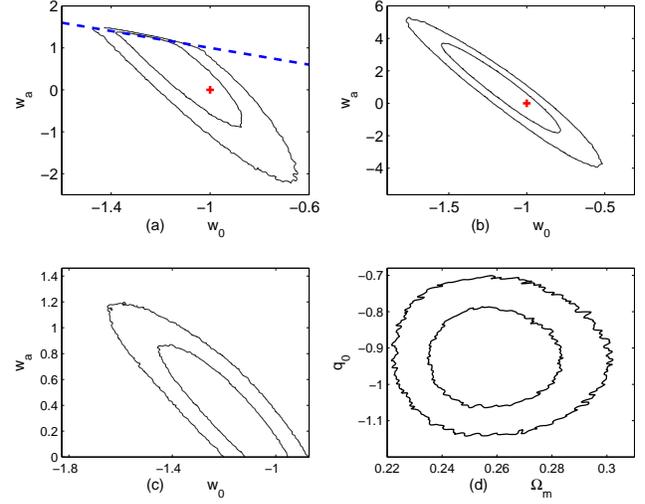}
\caption{The marginalized $1\sigma$ and $2\sigma$ contour plots of $w_0$ ($\Omega_m$) and $w_a$ ($q_0$)
for the CPL (a), JBP (b), Wetterich (c) and $q(z)$ (d) parametrisations. The dashed line in the upper left panel denotes
the condition $w_0+w_a=0$.}
\label{u2cont}
\end{figure}

\begin{figure}
\includegraphics[width=84mm]{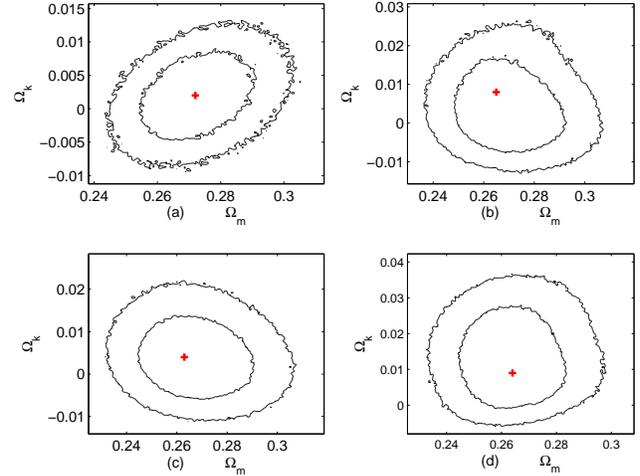}
\caption{The marginalized $1\sigma$ and $2\sigma$ contour plots of $\Omega_m$ and $\Omega_k$
for the $\Lambda$CDM (a), CPL (b), JBP (c) and Wetterich (d)  parametrisations. The red cross denotes the
best-fitting value.}
\label{u2omkcont}
\end{figure}

Now we consider the parametrization proposed by
\cite{wetterich},
\begin{equation}
\label{wet}
w(z)=\frac{w_0}{[1+w_a\ln(1+z)]^2}.
\end{equation}
For this model, $w(z=0)=w_0$ and $w(z)\sim 0$ when $z\gg 1$, so the behaviour
of $w(z)$ at high redshift is limited. The dark energy density is
\begin{equation}
\label{wetde}
\Omega_{DE}=(1-\Omega_{m}-\Omega_{k}-\Omega_{r})(1+z)^{3+3 w_0/[1+w_a\ln(1+z)]}.
\end{equation}
In this model, the model parameters are $\mathbf{p}=(\Omega_{m},\ \Omega_{k},  \ w_0, \ w_a)$.
Fitting the model to the combined SN Ia, Bao2, Baoz, WMAP7 and $H(z)$ data, we get the marginalized
$1\sigma$ constraints, $\Omega_m=0.264\pm 0.013$, $\Omega_k=0.009^{+0.014}_{-0.005}$,
$w_0=-1.17^{+0.09}_{-0.23}$, and $w_a=0.32^{+0.46}_{-0.16}$ with $\chi^2=540.4$.
The contours of $w_0$ and $w_a$
are plotted in Fig. \ref{u2cont}(c), and the contours of $\Omega_m$ and $\Omega_k$
are plotted in Fig. \ref{u2omkcont}(d).

\begin{figure}
\includegraphics[width=84mm]{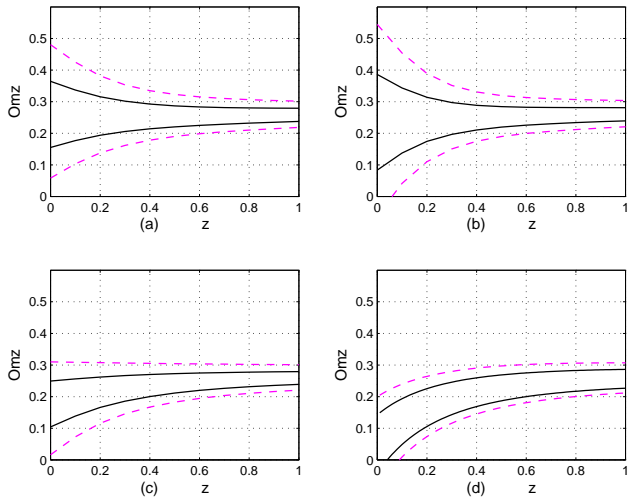}
\caption{The marginalized $1\sigma$ and $2\sigma$ errors of $Om(z)$ for the CPL (a), JBP (b), Wetterich (c) and $q(z)$ (d) parametrisations.}
\label{u2omz}
\end{figure}

For the flat Wetterich model, $Om(z)$ becomes
\begin{equation}
\label{wetomz}
Om(z)=\frac{\Omega_m (1+z)^3+\Omega_{DE}(z)-1}{(1+z)^3-1},
\end{equation}
where $\Omega_{DE}(z)$ is defined in equation (\ref{wetde}) with $\Omega_k=0$.
By fitting the combined data to the flat Wetterich model, we get the marginalized $1\sigma$ constraints,
$\Omega_m=0.266^{+0.01}_{-0.015}$, $w_0=-1.05^{+0.02}_{-0.16}$, and $w_a=0.14\pm 0.1$
with $\chi^2=541.1$. Using this result, we reconstruct $Om(z)$ with equation (\ref{wetomz}) and
the result is shown in Fig. \ref{u2omz}(c).

\subsection{Piecewise parametrization of $w(z)$}

\begin{figure}
\includegraphics[width=84mm]{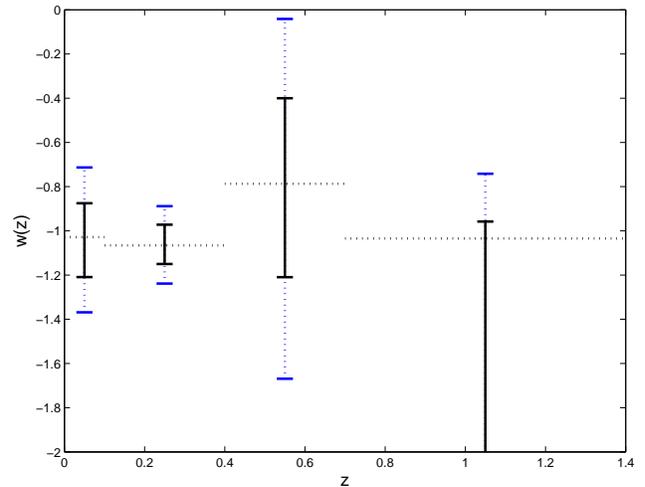}
\caption{The $1\sigma$ and $2\sigma$ estimates of the four uncorrelated parameters $\mathcal{W}_i$.}
\label{wzunc}
\end{figure}

Finally, we consider a more model-independent parametrization of $w(z)$, the piecewise parametrization of $w(z)$.
In this parametrization, the equation of state parameter is a constant, $w(z)=w_i$ for the redshift in the range $z_{i-1}<z<z_i$.
For convenience, we
choose $z_0=0$. We also assume that $w(z>1.8)=-1$. For a flat Universe, if $z_{i-1}\le z<z_i$,
\begin{equation}
\label{wcez}
\Omega_{DE}(z)=(1-\Omega_{m})(1+z)^{3(1+w_N)}\prod_{i=1}^{N}(1+z_{i-1})^{3(w_{i-1}-w_i)}.
\end{equation}
Again, the four parameters $w_i$ are correlated and we follow \cite{huterer05} to transform these parameters
to decorrelated parameters $\mathcal{W}_i$. By fitting the model to the combined SN Ia, Bao2, Baoz, WMAP7 and $H(z)$ data,
we get the error estimations of $\mathcal{W}_i$ and the results are shown in Fig. \ref{wzunc}.


\begin{figure}
\includegraphics[width=84mm]{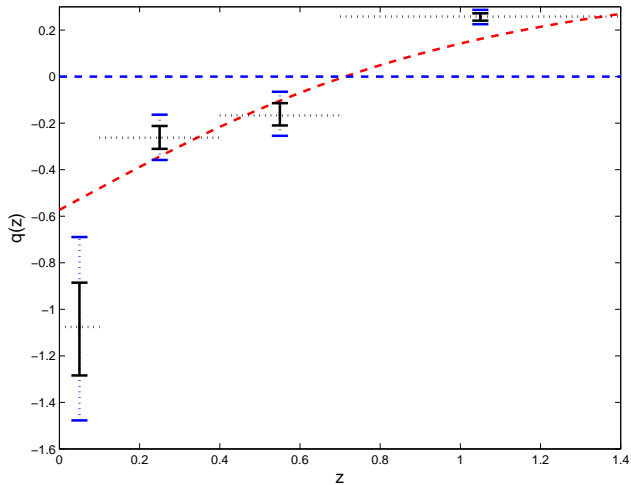}
\caption{The $1\sigma$ and $2\sigma$ errors of the four uncorrelated $\mathcal{Q}_i$.
The red dashed line is reconstructed with the best-fitting $\Lambda$CDM model.}
\label{qzunc}
\end{figure}

\begin{table*}
\label{table1}
\begin{minipage}{126mm}
 \caption{The marginalized $1\sigma$
errors for parameters constrained by the observational data}
\begin{tabular}{cccccccc}
\hline
 & $\chi^2$/DOF & $\Omega_{m}$& $\Omega_{k}$& $w_0\ (q_1)$& $w_a\ (q_2)$ & AIC &BIC \\ \hline
$\Lambda$CDM &541.2/576 & $0.272_{-0.011}^{+0.013}$ & $0.002\pm 0.004$ & & & 545.2& 553.9\\  \hline
DGP &561.6/576 & $0.288_{-0.011}^{+0.015}$  & $0.019\pm 0.005$ & & & 565.6& 574.3\\  \hline
CPL & 540.5/574 & $0.265^{+0.019}_{-0.009}$  & $0.008^{+0.005}_{-0.011}$ & $-1.16_{-0.06}^{+0.26}$ & $0.69^{+0.24}_{-1.41}$ & 548.5&565.9 \\  \hline
JBP  & 540.6/574 & $0.263^{+0.02}_{-0.01}$ & $0.004\pm 0.006$ & $-1.21_{-0.18}^{+0.32}$ & $1.29^{+1.35}_{-2.33}$ & 548.6& 566.0\\  \hline
Wetterich & 540.4/574 & $0.264\pm 0.013$  & $0.009_{-0.005}^{+0.014}$ & $-1.17^{+0.09}_{-0.23}$ & $0.32_{-0.16}^{+0.46}$ & 548.4& 565.8\\  \hline
$q_1-q_2$ model & 542.6/576 &  & & $0.07\pm 0.11$ & $-1.43\pm 0.09$ & 546.6& 555.3\\  \hline
flat CPL & 541.1/575 & $0.267^{+0.019}_{-0.01}$  &  & $-1.05_{-0.1}^{+0.17}$ & $0.07^{+0.32}_{-0.88}$ & 547.1& 560.2\\  \hline
flat JBP  & 541.0/575 & $0.265^{+0.019}_{-0.011}$ &  & $-1.08_{-0.19}^{+0.24}$ & $0.32^{+1.01}_{-1.72}$ & 547.0& 560.1 \\  \hline
flat Wetterich & 541.1/575 & $0.266_{-0.015}^{+0.01}$  &  & $-1.05^{+0.02}_{-0.16}$ & $0.14\pm 0.1$ & 547.1& 560.2\\  \hline
\end{tabular}
\end{minipage}
\end{table*}

\section{Conclusions}

We summarize all the results in Table 1 and some results are shown in Figs. \ref{u2cont}-\ref{wzunc}.
By parametrizing the deceleration parameter $q(z)$, we find very strong evidence for the current acceleration.
For the piecewise parametrization of $q(z)$, we find that $q(z)<0$ in the redshift range $0\le z\la 0.6$,
and $q(z)>0$ in the redshift range $z>0.8$ as shown in Fig. \ref{qzunc}. So the Universe experiences accelerated
expansion up to the redshift $z\sim 0.6$ and decelerated expansion at large redshift $z>0.8$.
For the CPL, JBP and Wetterich models, we see from Fig. \ref{u2cont} that $\Lambda$CDM model is consistent
with them, and this is further confirmed by the $Om$ diagnostic as shown in Fig. \ref{u2omz}. The
piecewise parametrization of $w(z)$ also confirms that $\Lambda$CDM model is consistent with current observations
as shown in Fig. \ref{wzunc}.
The CPL, JBP and Wetterich models differ
in the behaviour of $w(z)$ at high redshift; from Table 1 we see that all of them fit the observational data well,
and $w(z)\la 0$ as seen in Fig. \ref{u2cont}(a).
So the current data is still unable to distinguish models with different behaviours of $w(z)$ at a high redshift.
The number of parameters for $\Lambda$CDM and DGP models are the same, the difference between the best-fitting value of $\chi^2$
is $\Delta \chi^2=20.4$, so DGP model is excluded by the current data at more than $3\sigma$ level. The observational
constraint on the growth index $\gamma$ is $\gamma=0.46^{+0.12}_{-0.08}$ for the DGP model which is more than $1\sigma$
away from the theoretical value $11/16$, and the growth index of $\Lambda$CDM model is $\gamma=0.56^{+0.14}_{-0.09}$ which
is consistent with the theoretical value $6/11$.
Our results also show that the Universe is almost flat.
By using the prior $-5\le \log|\Omega_k|\le 0$, it was found that
$-0.9\times 10^{-2}\le \Omega_k\le 0.01$ at 99\% confidence level with the Bayesian model averaging method \citep{vardanyan11}.
In order to compare different models with different number of parameters,
we usually apply Akaike Information Criterion (AIC) \citep{aic}.
In terms of AIC, $\Lambda$CDM model is slightly preferred by the current data.
Furthermore, to account for the effects of the number of data points and the number of parameters,
Bayesian Information Criterion (BIC) \citep{bic} is used for model comparison. In terms of BIC,
the $\Lambda$CDM model is again preferred by the current data.
In addition to the approximate methods like AIC or BIC for model comparison, the Bayesian
model comparison provides a better tool for model selection \citep{trotta08}.

Our results are based on the standard $\chi^2$ method which has some shortcomings \citep{march},
so \cite{march} presented the Bayesian hierarchical method to constrain the cosmological parameters
and argued that the new method gives tighter constraint and outperforms the standard $\chi^2$ method.

\section*{acknowledgments}

This work was partially supported by the NNSF key project of China under grant No. 10935013,
the National Basic Science Program (Project 973) of China under
grant Nos. 2007CB815401 and 2010CB833004, CQ CSTC under
grant No. 2009BA4050 and CQ CMEC under grant No. KJTD201016. Gong thanks the hospitality of the Abdus Salam International Centre
of Theoretical Physics where the work was finished. Z-HZ was partially supported by the NNSF
Distinguished Young Scholar project under Grant No. 10825313.


\begin{thebibliography}{}
\bibitem[\protect\citeauthoryear{Akaike}{1974}]{aic} Akaike H., 1974, IEEE Trans. Auto. Control, 19, 716
\bibitem[\protect\citeauthoryear{Amanullah et al.}{2010}]{union2} Amanullah R. et al., 2010, ApJ, 716, 712
\bibitem[\protect\citeauthoryear{\^{A}ngela et al.}{2008}]{angela} da \^{A}ngela J. et al., 2008, MNRAS, 383, 565
\bibitem[\protect\citeauthoryear{Schwarz}{1974}]{bic} Schwarz G., 1978, Ann. Stat., 5, 461
\bibitem[\protect\citeauthoryear{Blake et al.}{2010}]{blake10} Blake C. et al., 2010, MNRAS, 406, 803
\bibitem[\protect\citeauthoryear{Cai, Su \& Zhang}{2010}]{cai} Cai R. G., Su Q. P.,  Zhang H.-B., 2010, J. Cosm. Astropart. Phys., 04, 012
\bibitem[\protect\citeauthoryear{Chevallier \& Polarski}{2001}]{cpl1} Chevallier M.,  Polarski D., 2001, Int. J. Mod.
Phys. D, 10, 213
\bibitem[\protect\citeauthoryear{Clarkson, Cort\^{e}s \& Bassett}{2007}]{clarkson} Clarkson C., Cort\^{e}s M.,
Bassett B., 2007, J. Cosm. Astropart. Phys., 08, 011
\bibitem[\protect\citeauthoryear{Conley et al.}{2008}]{conley} Conley A. et al., 2008, ApJ, 681, 482
\bibitem[\protect\citeauthoryear{Dossett et al.}{2010}]{ishak} Dossett J., Ishak M., Moldenhauer J., Gong Y.G., Wang A., 2010, J. Cosm. Astropart. Phys., 04, 022
\bibitem[\protect\citeauthoryear{Dvali, Gabadadze \& Porrati}{2000}]{dgp} Dvali G., Gabadadze G., Porrati, M., 2000,  Phys. Lett. B, 485, 208
\bibitem[\protect\citeauthoryear{Eisenstein \& Hu}{1998}]{dw} Eisenstein D. J., Hu W., 1998, ApJ, 496, 605
\bibitem[\protect\citeauthoryear{Eisenstein et al.}{2005}]{baoa} Eisenstein D. J. et al., 2005, ApJ, 633, 560
\bibitem[\protect\citeauthoryear{Gazta\~{n}aga, Miquel \& S\'{a}nchez}{2009a}]{eg} Gazta\~{n}aga E., Miquel, R., S\'{a}nchez E., 2009a, Phys. Rev. Lett., 103, 091302
\bibitem[\protect\citeauthoryear{Gazta\~{n}aga, Cabr\'{e} \& Hui}{2009b}]{hz2} Gazta\~{n}aga E., Cabr\'{e} A., Hui L., 2009b,  MNRAS, 399, 1663
\bibitem[\protect\citeauthoryear{Gong \& Wang}{2007}]{gong07a} Gong Y. G., Wang, A., 2007, Phys. Rev. D, 75, 043520
\bibitem[\protect\citeauthoryear{Gong}{2008}]{gongprd08} Gong Y.G., 2008, Phys. Rev. D, 78, 123010
\bibitem[\protect\citeauthoryear{Gong, Wu \& Wang}{2008}]{gong08} Gong Y. G., Wu Q., Wang A., 2008, ApJ, 681, 27
\bibitem[\protect\citeauthoryear{Gong, Ishak \& Wang}{2009}]{gong09} Gong Y. G., Ishak M., Wang A., 2009, Phys. Rev. D, 80, 023002
\bibitem[\protect\citeauthoryear{Gong et al.}{2010a}]{gong10a} Gong Y. G., Cai R. G., Chen Y, Zhu Z. H., 2010a, J. Cosm. Astropart. Phys., 01, 019
\bibitem[\protect\citeauthoryear{Gong, Wang \& Cai}{2010b}]{gong10b} Gong Y. G., Wang B., Cai R.G., 2010b, J. Cosm. Astropart. Phys., 04, 019
\bibitem[\protect\citeauthoryear{Guzzo et al.}{2008}]{guzzo} Guzzo L. et al., 2008, Nature, 451, 541
\bibitem[\protect\citeauthoryear{Hicken et al.}{2009}]{consta} Hicken M., Wood-Vasey W. M., Blondin S.,
Challis P., Jha S., Kelly P. L., Rest A., Kirshner R. P., 2009, ApJ, 700, 1097
\bibitem[\protect\citeauthoryear{Hu \& Sugiyama}{1996}]{hs} Hu W., Sugiyama N., 1996, ApJ, 471, 542
\bibitem[\protect\citeauthoryear{Huang et al.}{2009}]{huang} Huang Q. G., Li M., Li X. D., Wang S., 2009, Phys. Rev. D, 80, 083515
\bibitem[\protect\citeauthoryear{Huterer \& Cooray}{2005}]{huterer05} Huterer D., Cooray A., 2005, Phys. Rev. D, 71, 023506
\bibitem[\protect\citeauthoryear{Jassal, Bagla \& Padmanabhan}{2005}]{jbp} Jassal H. K., Bagla J. S., Padmanabhan T., 2005, MNRAS, 356, L11
\bibitem[\protect\citeauthoryear{Kessler et al.}{2010}]{sdss2} Kessler R. et al., 2010, ApJS, 185, 32
\bibitem[\protect\citeauthoryear{Komatsu et al.}{2011}]{wmap7} Komatsu E. et al., 2011, ApJS, 192, 18
\bibitem[\protect\citeauthoryear{Kowalski et al.}{2008}]{union} Kowalski M. et al., 2008, ApJ, 686, 749
\bibitem[\protect\citeauthoryear{Lampeitl et al.}{2009}]{lampeitl} Lampeitl H. et al., 2009, MNRAS, 401, 2331
\bibitem[\protect\citeauthoryear{Lewis \& Bridle}{2002}]{cosmomc} Lewis A., Bridle S., 2002, Phys. Rev. D, 66, 103511
\bibitem[\protect\citeauthoryear{Linder}{2003}]{cpl2} Linder E. V., 2003, Phys. Rev. Lett., 90, 091301
\bibitem[\protect\citeauthoryear{March et al.}{2011}]{march} March M. C., Trotta R., Berkes P., Starkman G. D.,
Vaudrevange P. M., 2011, arXiv: 1102.3237
\bibitem[\protect\citeauthoryear{McDonald et al.}{2005}]{mcdonald} McDonald P. et al., 2005, ApJ, 635, 761
\bibitem[\protect\citeauthoryear{Pan et al.}{2010}]{pan} Pan N.N., Gong Y.G., Chen Y., Zhu Z.H., 2010, Class. Quantum Grav., 27, 155015
\bibitem[\protect\citeauthoryear{Percival et al.}{2007}]{percival07} Percival W. J., Cole S., Eisenstein D. J.,
Nichol R. C., Peacock J. A., Pope A. C., Szalay A. S., 2007,  MNRAS, 381, 1053
\bibitem[\protect\citeauthoryear{Percival et al.}{2010}]{wjp} Percival W. J. et al., 2010,  MNRAS, 401, 2148
\bibitem[\protect\citeauthoryear{Perlmutter et al.}{1999}]{acc2} Perlmutter S. et al., 1999, ApJ, 517, 565
\bibitem[\protect\citeauthoryear{Riess et al.}{1998}]{acc1} Riess A. G. et al., 1998, AJ,  116, 1009
\bibitem[\protect\citeauthoryear{Riess et al.}{2009}]{riess09}  Riess A. G. et al., 2009, ApJ, 699, 539
\bibitem[\protect\citeauthoryear{Ross et al.}{2007}]{ross} Ross N.P. et al., 2007, MNRAS, 381, 573
\bibitem[\protect\citeauthoryear{Sahni, Shafieloo \& Starobinsky}{2008}]{omz} Sahni V., Shafieloo A., Starobinsky A. A., 2008, Phys. Rev. D, 78, 103502
\bibitem[\protect\citeauthoryear{Serra et al.}{2009}]{corray9} Serra  P., Cooray A., Holz D. E.,
Melchiorri A., Pandolfi S., Sarkar D., 2009, Phys. Rev. D, 80, 121302
\bibitem[\protect\citeauthoryear{Shafieloo, Sahni \& Starobinsky}{2009}]{star} Shafieloo A., Sahni V., Starobinsky, A. A., 2009, Phys. Rev. D, 80, 101301
\bibitem[\protect\citeauthoryear{Sollerman et al.}{2009}]{sollerman} Sollerman J. et al., 2009, ApJ, 703, 1374
\bibitem[\protect\citeauthoryear{Stern et al.}{2010}]{hz1} Stern D., Jimenez R., Verde L., Kamionkowski M.,
Stanford S. A., 2010, J. Cosm. Astropart. Phys., 02, 008
\bibitem[\protect\citeauthoryear{Tegmark el al.}{2006}]{tegmark} Tegmark M. el al., 2006, Phys. Rev. D, 74, 123507
\bibitem[\protect\citeauthoryear{Trotta}{2008}]{trotta08} Trotta R., 2008, Contemporary Phys., 49, 71
\bibitem[\protect\citeauthoryear{Vardanyan, Trotta \& Silk}{2011}]{vardanyan11} Vardanyan M., Trotta R., Silk J., 2011, MNRAS, 413, L91.
\bibitem[\protect\citeauthoryear{Viel, Haehnelt \& Springel}{2004}]{viel1} Viel M., Haehnelt M.G., Springel V., 2004, MNRAS, 354, 684
\bibitem[\protect\citeauthoryear{Viel, Haehnelt \& Springel}{2006}]{viel2} Viel M., Haehnelt M.G., Springel V., 2006, MNRAS, 365, 231
\bibitem[\protect\citeauthoryear{Wetterich}{2004}]{wetterich} Wetterich C., 2004, Phys. Lett. B, 594, 17
\end{thebibliography}
\end{document}